\documentclass[printer]{aa} 
\usepackage{graphicx}
\begin{document}
\title{Long-lived, long-period radial velocity variations in Aldebaran:  \\
A planetary companion and stellar activity}


   \author{A. P. Hatzes\inst{1}
          \and
   W. D. Cochran\inst{2}
          \and
    M. Endl\inst{2}
\and
   E. W. Guenther\inst{1}
	 \and
   P. MacQueen \inst{2}
\and
	M. Hartmann\inst{1}
          \and
\\
M. Zechmeister\inst{3}
	 \and
	I. Han\inst{4}
         \and
	B.-C. Lee\inst{4}
         \and
	G.A.H. Walker\inst{5}
	 \and
	S. Yang\inst{6} 
         \and
        A.M. Larson\inst{7}
         \and
             K.-M. Kim\inst{4}
         \and
\\
	D. E.  Mkrtichian\inst{8}$^,$\inst{9}
\and
   M. D\"ollinger\inst{1}
\and 
   A.E.  Simon\inst{10}$^,$\inst{11}
\and
L. Girardi\inst{12}
         }

   \offprints{
    Artie Hatzes, \email{artie@tls-tautenburg.de}\\$*$~
    Based in part on observations obtained at the
    2-m-Alfred Jensch Telescope at the Th\"uringer
    Landessternwarte Tautenburg and the telescope facilities 
   of McDonald Observatory
  \\}

     \institute{Th\"uringer Landessternwarte Tautenburg,
                Sternwarte 5, D-07778 Tautenburg, Germany
\and
	McDonald Observatory, The University of Texas at Austin,
    Austin, TX 78712, USA
\and 
Institut f\"ur Astrophysik, Georg-August-Universit\"at, Friedrich-Hund-Platz 1, 37077, G\"ottingen, Germany 
\and
 Korea Astronomy and Space Science Institute, 776, Daedeokdae-Ro, 
Youseong-Gu, Daejeon 305-348, Korea
\and
1234 Hewlett Place, Victoria, BC, V8S 4P7, Canada
\and
Department of Physics and Astronomy, University of Victoria,
Victoria, BC, V8W 3P6, Canada
\and
Astronomy Department, Box 351580, University of Washington, Seattle, WA 98195-1580
\and
National Astronomical Research Institute of Thailand, 191 Siriphanich Bldg.,
Huay Kaew Rd., Suthep, Muang, 50200 Chiang Mai, Thailand
\and
Crimean Astrophysical Observatory, Nauchny, Crimea, 98409, Ukraine
\and
Physikalisches Institut, Center for Space and Habitability, University of
Bern, Sidlerstrasse 5, CH-3012 Bern, Swithzerland
\and
 Konkoly Observatory, Research Centre for Astronomy and Earth Sciences,
Hungarian Academy of Sciences, H-1121 Budapest, Konkoly Thege. Mikl\'os.
\'ut 15-17, Hungary
\and
INAF-Osservatorio Astronomico di Padova, Vicolo dell'O servatorio 5, I-35122
Padova, Italy
}

   \date{Received; accepted}

 
  \abstract
{}
   {We investigate the nature of the long-period radial velocity
variations in $\alpha$ Tau first reported over 20 years ago.
   }
   {We analyzed  precise stellar radial velocity measurements
for $\alpha$ Tau spanning over 30 years.
An examination of the H$\alpha$ and Ca II $\lambda$8662 spectral lines,
and Hipparcos photometry  was  also done to help discern
the nature of the long-period radial velocity variations.
   }
   {
Our radial velocity  data show that the
long-period, low amplitude radial velocity  variations
are
long-lived and coherent. Furthermore, H$\alpha$ equivalent width  measurements
and Hipparcos photometry  show no significant
variations with this period. 
Another investigation of this star 
established that there was no variability  in the
spectral line shapes with the radial velocity period.
An orbital solution results in a period of $P$ = 628.96 $\pm$ 0.90 d,
eccentricity, $e$ = 0.10 $\pm$ 0.05, and a radial velocity amplitude,
$K$ = 142.1 $\pm$ 7.2 m\,s$^{-1}$. Evolutionary tracks yield a 
stellar mass of 1.13 $\pm$ 0.11 $M_\odot$, which corresponds to a minimum companion
mass of 6.47 $\pm$ 0.53 M$_{Jup}$ with an orbital semi-major axis of $a$ = 1.46 $\pm$ 0.27 AU.
After removing the orbital motion of the companion, an additional period
of $\approx$ 520 d is found in the radial velocity data, but only in some time
spans.
A similar period
is found in the variations in the equivalent width of H$\alpha$ and Ca II.
Variations at one-third of this period are also found in the spectral line bisector measurements.
The $\sim$ 520\,d period is interpreted as the rotation modulation by stellar surface structure.
{ Its presence, however, may not be long-lived,  and it only appears 
in epochs of the
radial velocity data separated by $\sim$ 10 years. This might be due to an activity
cycle.}
}
   {The  data presented here provide further evidence of  a 
planetary companion to $\alpha$ Tau,  as well as activity-related radial 
velocity variations.}

\titlerunning{Further evidence of a planetary companion to $\alpha$ Tau}
\maketitle

%

\section{Introduction}

Low amplitude radial
velocity (RV) variations with a period of 645 days
were first reported in the K giant star $\alpha$ Tau
($=$ Aldebaran $=$ HR 1457 $=$ HD 29139 $=$ HIP 21421) by 
Hatzes \& Cochran 1993 (hereafter HC93). HC93 hypothesized
that one cause for these variations was a planetary companion with
$m$ sin $i$ = 11.4  $M_{Jup}$, when assuming a stellar mass of 2.5
$M_\odot$.
There was some hint that these RV variations 
were long-lived because measurements  taken almost a decade earlier
by Walker et al. (1989) were consistent in amplitude
and phase with the measurements of HC93.
A subsequent analysis of the spectral line shapes (line bisectors)
by Hatzes \& Cochran
(1998) showed no evidence of variations {with the 645\,d RV period},
 although they did find evidence of  a 50\,d period
in the line bisectors that they attributed to stellar oscillations.
This seemed to provide further support for the presence of
a planetary companion. However, the confirmation of such a hypothesis 
was still in doubt owing to the unknown intrinsic variability of K giant
stars. 

\begin{figure}[h]
\resizebox{\hsize}{!}{\includegraphics{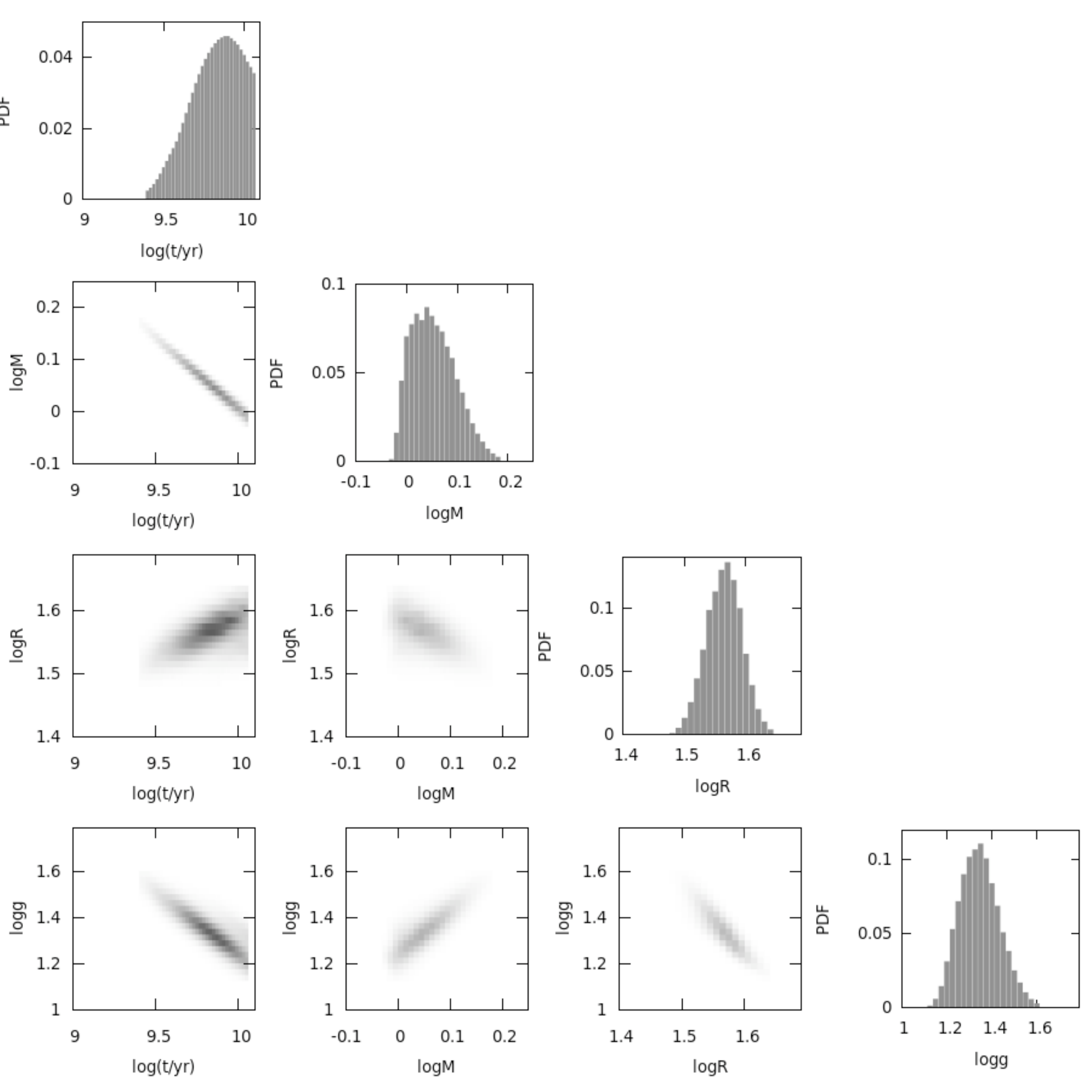}}
\caption{The probability density functions (PDF) for various parameters
of $\alpha$ Tau.  The histogram (on right) for each row shows the PDF for
(top to bottom) stellar age, mass, radius, and surface gravity. The correlation
plots on each row show how that parameter correlates with
the other parameters. 
}
\label{PDF}
\end{figure}

Subsequent to the discovery of RV variations in Aldebaran it has been
well established that K giant stars can indeed possess giant extrasolar
planets (e.g., Frink et al. 2002; Setiawan et al. 2003;
Hatzes et al. 2005; D\"ollinger et al. 2007; Niedzielski et al. 2009).
In particular, the long period RV variations for 
one of the K giants studied by HC93, $\beta$ Gem, were recently confirmed to be
caused by a giant planet with mass of 2.3 $M_{Jup}$ (Hatzes et al. 2006).
 This was  
established by long-lived, coherent RV variations spanning 25 years,
along with the lack of variations in the spectral line shapes and Ca II
emission. Reffert et al. (2006) and Han et al. (2008)
also confirmed the long-lived nature of the RV
variations of $\beta$ Gem. 
{ RV variations caused by a substellar companion should not show any 
variations in the spectral lines with the same period as the planet.
In the case of $\alpha$~Tau, this has already been
established by the line bisector measurements of Hatzes \& Cochran (1998).
Long-lived RV variations lasting several decades would provide 
additional compelling evidence for the planet hypothesis, as was the case
for $\beta$ Gem.} For $\alpha$ Tau
this is problematical  as this star can show intrinsic variations
of $\approx$ 100 m\,s$^{-1}$.
A large number of measurements over a long
time base are needed to extract the long period RV signal with 
confidence.

	Here, we present additional measurements of $\alpha$ Tau taken
with the coude {\'e}chelle spectrograph of the  Th\"uringer Landessternwarte
Tautenburg (TLS), the Tull Spectrograph of the
McDonald 2.7m telescope, and the Bohyunsan Observatory {\'E}chelle Spectrograph
(BOES) spectrograph of the Bohyunsan Optical Astronomy
Observatory (BOAO). These measurements are  combined with previous measurements
to produce a total baseline of over 30 years. The combined data 
clearly show that a period near 629 days has been present with the same amplitude and phase
for the entire time span lending further evidence to the planet hypothesis
for the long period RV variations of $\alpha$ Tau. These data also show
additional variations which are most likely due to rotational modulation
by surface features.

\section{Stellar parameters }

The K5 III star $\alpha$ Tau is at a distance of 
20.43 $\pm$ 0.32 pc as measured by Hipparcos
(van Leeuwen 2007).  Richichi \& Roccatagliata (2005) presented
a few very accurate radius measurements of $\alpha$ Tau.
They determined
a limb-darkened  angular diameter
of 20.058 $\pm$ 0.03 mas and a uniform disk angular diameter of
19.96 $\pm$ 0.03 mas yielding
a radius of 45.1 $\pm$ 0.1 $R_\odot$ for the limb-darkened
case. By virtue of its large angular diameter and location in 
the ecliptic there
have been a large number of angular diameter measurements for $\alpha$
Tau using 
interferometry or  lunar occultations.
Richichi \& Roccatagliata (2005) discussed these in detail
and noted that there
was considerable scatter in these measurements corresponding to 
$R$ = 36 -- 46 $R_\odot$ at 550 nm
that was larger than the formal errors. The source of such discrepancies
is not known.

\begin{figure*}
\resizebox{\hsize}{!}{\includegraphics{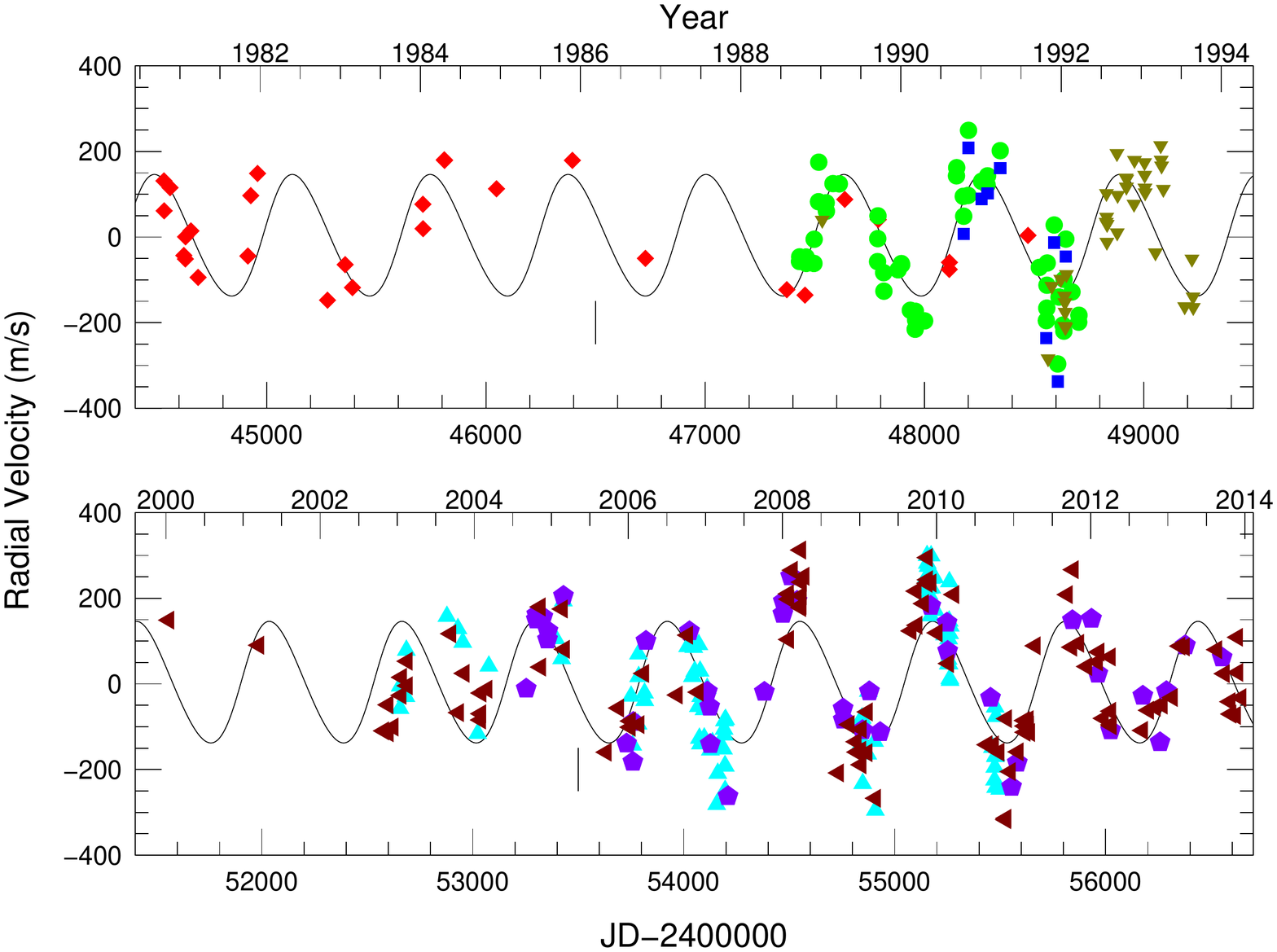}}
\caption{Radial velocity measurements for $\alpha$ Tau from the seven  data sets:
CFHT (diamonds), DAO (inverted triangles), McD-2.1m (circles),
 McD-CS11 (squares), McD-Tull (sideways triangles),
TLS (triangles), and BOAO (pentagons). Zero point corrections have
been applied to the individual data sets before plotting (see text).
The curve represents the 
orbital solution (see Table 10). The vertical dash represents the 
peak-to-peak variations of the stellar oscillations. 
}
\label{orbit}
\end{figure*}

\begin{figure}
\resizebox{\hsize}{!}{\includegraphics{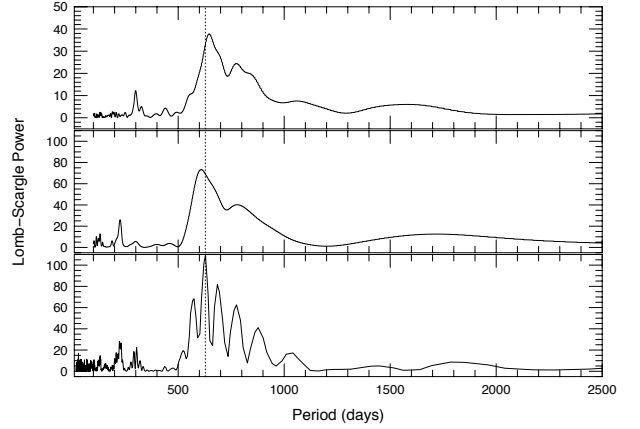}}
\caption{L-S periodogram of the RV measurements taken prior
to JD =  2450000 (top), after JD = 2450000 (middle), and the full data set
(bottom).
}
\label{periodogram}
\end{figure}

\begin{figure}
\resizebox{\hsize}{!}{\includegraphics{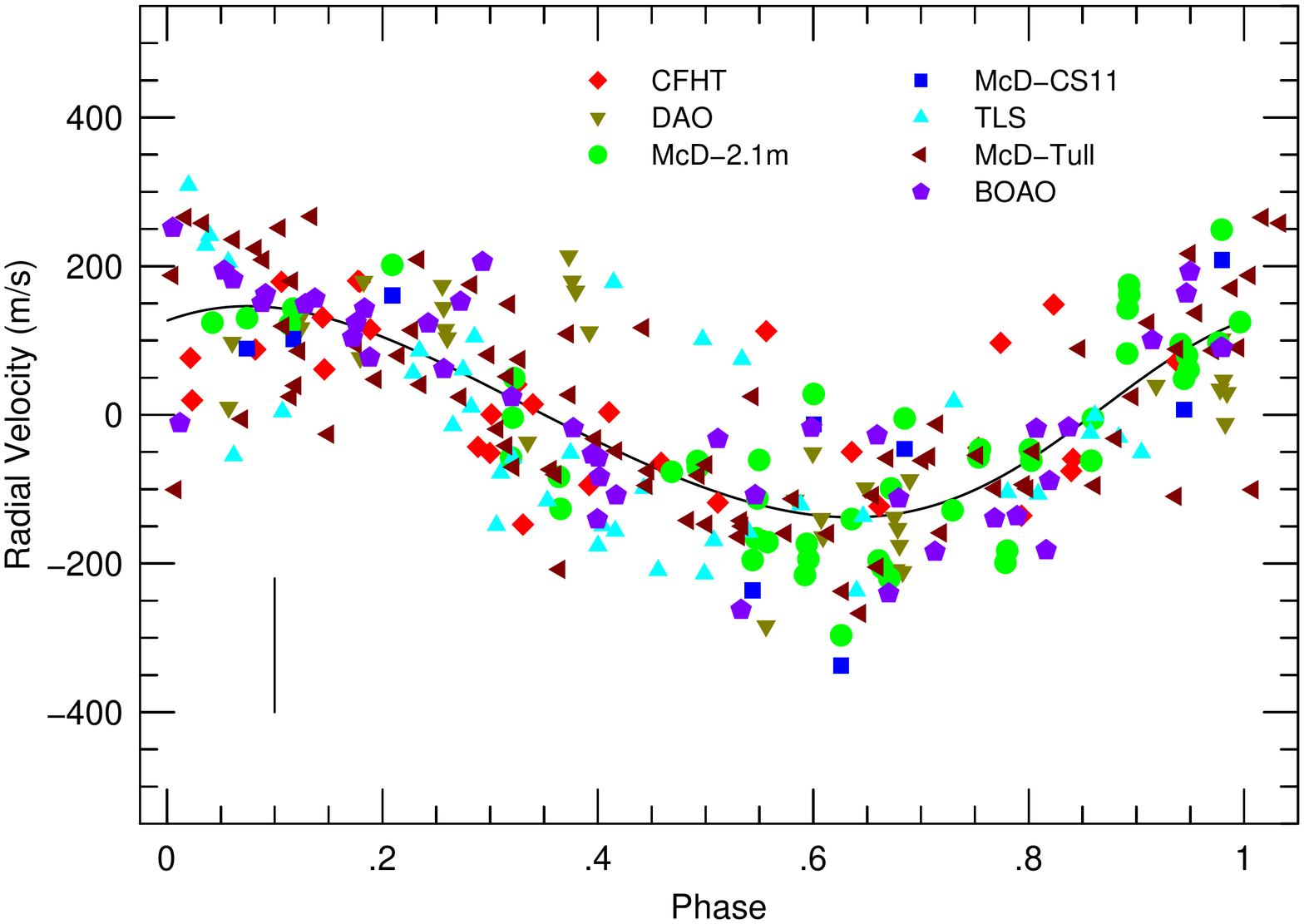}}
\caption{Radial velocity measurements for $\alpha$ Tau from the seven data sets phased to the
orbital period. { For the TLS and McD-Tull data we plot 
run-binned averages that were used in the orbital solution.}
The solid line is the orbital solution and the vertical line the
peak-peak short term intrinsic RV variations observed for this star. 
Symbols are the same as for Fig.~\ref{orbit}. Error bars for individual 
measurements are not shown for clarity and because the
scatter is dominated by the intrinsic stellar variability. 
}
\label{orbitphase}
\end{figure}

McWilliam (1990)
measured an effective temperature of 3910 K, a metalicity of 
$[Fe/H]$ $=$ $-$0.34 $\pm$ 0.21,  and a surface gravity, log $g$ $=$ 1.59 $\pm$ 0.27.
We also derived the stellar parameters for $\alpha$ Tau using a high
signal-to-noise  spectrum taken with the coude {\'e}chelle spectrograph of the
TLS. {The details of the analysis method will be given in a forthcoming paper.
Briefly, we used the model atmospheres  of Gustafsson et al. (1975)
and plane-parallel geometry along with the
assumption of local thermodynamic equilibrium. The iron content
was determined from the equivalent width of over one hundred Fe I lines and eight
Fe II lines. 

The effective temperature was determined by imposing that the Fe I abundance
does not depend on the excitation potential of the lines. 
The surface gravity was determined
using the Fe I and Fe II ionization equilibrium.

Our analysis yielded $T_{\mathrm{eff}}$ = 4055 $\pm$ 70 K, log~g = 1.20 $\pm$ 0.1, and [Fe/H] = $-$0.27 
$\pm$ 0.05. Our abundance and log g values are consistent with the McWilliam (1990) analysis
and the value determined from the evolutionary tracks (Fig. 1).
The determination of $T_{\mathrm{eff}}$ is in line with the McWilliam result as
well as 
literature values from the past 20 years. These range  as low as
3875 K (Luck \& Challener 1995) and as high as 4131 K (Kovacs 1983). }

The basic stellar parameters such as mass, radius, and age were determined
using theoretical isochrones and a modified version of
Jo{$\!\!\!/$}rgensen {\&} Lindegren's (2005) method also described in 
da Silva et al. (2006). Figure~\ref{PDF} shows 
the probability density functions
 (PDF) of the derived parameters. 
The PDF for the age appears
truncated because the prior includes a maximum age for the star.
 The PDF mean values
and the 68\% confidence level for the { stellar age and mass}  are listed in Table 1.
 The first figures of each row also show how the current parameter
 (row) correlates with the previous (rows from above) parameters. For instance, a
radius $\sim$ 5\% larger than the one estimated  would imply a smaller log\,g by
$\sim$ 0.02 dex, a decrease of $\sim$ 4\% in mass, and an increase of  $\sim$ 28\% in age.

The radius determined by this method, $R$ $=$ 36.68 $\pm$ 2.46 $R_\odot$, is
less than the Richichi \& Roccatagliata (2005)
nominal value. However,  it is still within the range of 
radius measurements that have been made for this star. In Table 1 we list the interferometric
value of the radius as determined by Richichi \& Roccatagliata (2005).
The stellar mass derived by this method is 1.13 $\pm$ 0.11 $M_\odot$ which
we adopt for the subsequent analysis. 
{We note that the surface gravity determined
via our spectral analysis is consistent with the mass and radius for the star listed
in Table 1. 
Assuming that $\alpha$ Tau  did not lose
significant mass loss as it evolved to its present state,  then its progenitor
star on the main sequence would have had a spectral type of $\sim$ F7. }

\section{The radial velocity data}

Seven independent data sets of high precision radial velocity data were used
for our analysis.  Precise RV measurements were also made with a Hydrogen-Fluoride
(H-F) cell as part of the CFHT survey (hereafter the ``CFHT'' data set) of 
Walker et al. (1989) as well as additional measurements from the Dominion
Astrophysical Observatory (hereafter ``DAO'' data set) using the same
technique. See Campbell \& Walker  (1979) and
Larson et al. (1993b) for a description of the H-F measurements. For the
remaining
five RV data sets an iodine (I$_2$) cell provided the wavelength reference.
These include the original measurements using the McDonald Observatory
2.1m telescope (hereafter ``McD-2.1m'' data set) and
the coude spectrograph in the so-called  ``cs11'' focus (hereafter ``McD-CS11'' data set) 
of the 2.7m telescope at McDonald
Observatory. { We should note that the we did not include the McDonald data that were used
for the bisector measurements of Hatzes \& Cochran (1998). These were taken using
telluric lines as a wavelength reference which had a lower precision than the iodine
wavelength calibration or H-F methods.}

The latest McDonald measurements were taken using the
Tull Spectrograph at the  so-called ``cs23'' focus (hereafter the ``McD-Tull''
data set) as part of a long-term planet search program
(e.g. Cochran et al. 1997; Endl et al. 2004; Robertson et al. 2012).

Observations of $\alpha$ Tau  were made as part of the 
planet search program of the Th\"uringer Landessternwarte
Tautenburg (``TLS'' data set)
and at the Bohyunsan
Optical Astronomy Observatory. The TLS  program uses 
the
high resolution coude {\'e}chelle spectrometer  of the Alfred Jensch 2m
telescope and an iodine absorption cell placed { before the entrance slit
of the spectrograph.}
The instrument  is a grism crossed-dispersed  {\'e}chelle spectrometer that has
a resolving power $R$ ($\lambda$/$\Delta \lambda$) = 67,000 and 
wavelength coverage 4630--7370\,{\AA} when using the so-called ``Visual''
grism. A more detailed description of radial velocity measurements
from the TLS program can be found in Hatzes et al. (2005). 


The BOAO measurements (``BOAO'' data set) were made with the fiber-fed Bohyunsan Observatory
E\'chelle Spectrograph (BOES). BOES has a wavelength coverage of 
3600 -- 10500 {\AA}. The resolving power
for the observations was 90,000. An iodine absorption cell 
was also used to provide
the wavelength reference for the precise RV measurements. See Kim et al. 
(2006) for a more detailed description of the instrument and data analysis
used for these RV measurements.

Table 2  lists the journal of
observations which includes the data set, time coverage, the wavelength reference
employed, the number of observations, and the 
rms scatter about
the orbital solution presented in Section 4.

Tables 3--9 list
the RV measurements from the  seven data sets. 
Each instrument only 
measures a relative RV, thus each data set
has a different zero-point offset. In Section \S4  the relative
zero-point offsets were calculated as part of the global fit
to the orbital solution.
These offsets have been applied in the tabulated data and the
displayed measurements.
Alpha Tau is a bright star and typically multiple measurements were made
each night. Nightly averages were used in the analysis and the plots
and are listed in the
tables.

\begin{figure}
\resizebox{\hsize}{!}{\includegraphics{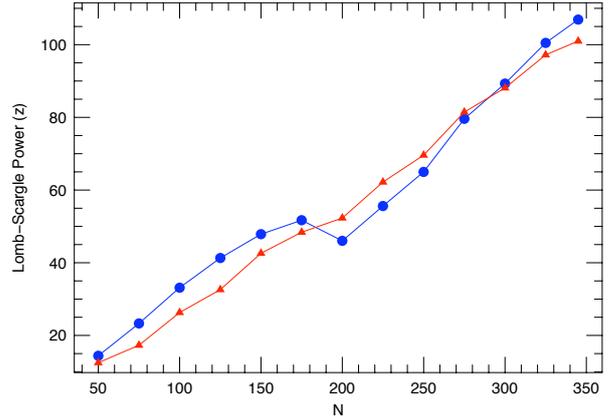}}
\caption{L-S power ($z$) of the 629\,d period
as a function of the number of data points
used in the periodogram calculation (circles). The power, $z$,
as a function of the number of data points for simulated data
(triangles).
}
\label{fapgrow}
\end{figure}


\begin{figure}[t]
\resizebox{\hsize}{!}{\includegraphics{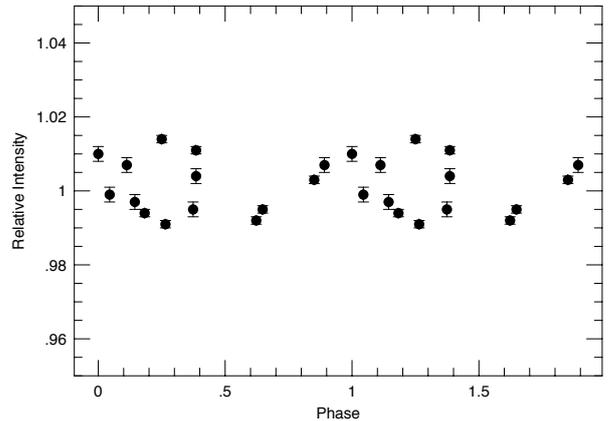}}
\caption{The Hipparcos photometry phased to the 629\,d orbital period.}
\label{photphase}
\end{figure}

The RV measurements for all data sets (with the appropriate offsets applied, see below)
are shown in Figure~\ref{orbit}.
There appears to be long-lived sinusoidal 
variations that span the entire data sets and have a scatter
of $\approx$ 100 m\,s$^{-1}$.

The bottom  panel of Figure~\ref{periodogram} shows the  Lomb-Scargle (L-S)
periodogram  (Lomb 1976; Scargle 1982) of all the RV measurements { with the
proper zero
point offsets as found in the orbit fitting subtracted from each data set (see below).}
Although the different data sets have a different precision, the
intrinsic RV jitter of the star is many factors larger than these 
and is approximately the same for all data sets. It is
therefore not important to weight the periodogram by the different errors.
There is a highly significant peak at a period, $P$ 
$\approx$ 630 days.  The false alarm probability (FAP) of this signal using
the equation in Scargle (1982)
is FAP $\approx$ 10$^{-20}$. This signal appears to be long-lived as it appears
in the RV measurements taken prior to 2000 (JD $<$ 2450000, top panel),
and after 2000 (middle panel).

\section{Orbital solutions}

An orbital solution was calculated using the  combined data sets and the
program {\it Gaussfit} (Jefferys et al. 1988). 
The velocity zero point for each
data set was allowed to be a free parameter. 
These individual zero points in 
the velocity were subtracted from each data set before plotting in 
Fig.~\ref{orbit}. (Tables 3--9
list the zero-point subtracted data.) 

Our initial solution resulted in a period of 
$P$ = 628.07 $\pm$ 0.82 d,  eccentricity, 
$e$ = 0.15 $\pm$ 0.04 and a velocity amplitude, $K$ = 158.0 $\pm$ 6.5 m\,s$^{-1}$.
{ The referee noted that some of the data sets (TLS and McD-Tull) had many 
measurements that were clumped in some epochs. This was particularly true
for the TLS data as numerous RV measurements were taken over several consecutive nights
in order to study the stellar oscillations. The referee expressed
a valid concern that this ``clumping'' may have given a higher
weight when computing the orbit. This could have a strong
effect primarily on the orbital eccentricity and velocity amplitude.}

{ Therefore, we binned the RV values over individual observing
 runs for the TLS and McD-Tull
data. This resulted in an average RV value over typically three to five nights. 
The orbital parameters using all the data, but the 
run-binned values from TLS and McD-Tull are listed in Table 10. 
These include the orbital period, $P$,  the time of periastron, the radial
velocity ($K$) amplitude, the eccentricity, $e$, the argument
of periastron, $\omega$, the mass function ($f(m)$),
the minimum companion mass, $m$ sin $i$, and the orbital semi-major
axis, $a$.  Note that using the run-binned
averages only results in a slight and insignificant 
decrease in the   eccentricity and $K$-amplitude.
Figure~\ref{orbitphase} shows
the RV measurements phase-folded and the orbital 
solution. Using run-binned averages results in a 
slightly, but not significantly lower eccentricity and $K$-amplitude.

\begin{figure}[h]
\resizebox{\hsize}{!}{\includegraphics{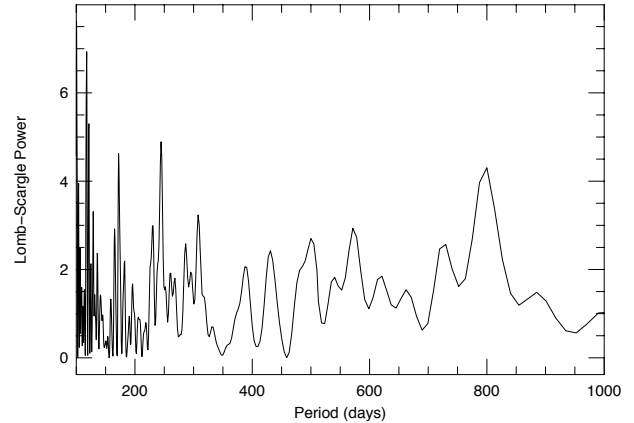}}
\caption{L-S periodogram of the $\Delta$EW$_{8662}$ variations.
}
\label{ftca}
\end{figure}

The rms scatter about the orbital solution
from the individual  data sets is  listed in Table 2
{and is about $\approx$ 100 m\,s$^{-1}$ for each set.}
This
is mostly due to intrinsic variability from stellar
oscillations. Alpha  Tau shows RV variations of several hundred
m\,s$^{-1}$
peak-to-peak in the course of several nights. For example, 
the RV of $\alpha$~Tau changed by { $\approx$ 220 m\,s$^{-1}$ between
JD = 2454071 and 2454076.}   This time scale and amplitude are typical
for oscillations in more evolved K giants (e.g. Hatzes \& Cochran 1994).
Therefore, we take $\approx$ $\pm$100 m\,s$^{-1}$
as the intrinsic RV ``jitter'' of the star and as our ``working'' precision.
This is shown by the vertical lines in Fig.~\ref{orbit} and Fig.~\ref{orbitphase}.

\section{The nature of the RV variations}

The fact that the RV variations seem to be long-lived and coherent
for over 30 years strongly argues that they are indeed due to a sub-stellar
companion. However, because $\alpha$~Tau is an evolved giant star it is prudent
to examine whether the RV period is present in other measured quantities.
As noted earlier,
Hatzes \& Cochran (1998) already established that there were no spectral line shape
variations  with a period of $\approx$ 645 d based on 
very high resolution ($R$ = 220,000) spectral data. 
The spectral line shape
measurements thus do not support rotational modulation as a cause for the 629\,d period RV
variations. Additional evidence can be found by examining the photometry and activity indicators as well as 
the stability of the RV period.

\subsection{Stability of the RV variations}

Although the orbital solution fits most of the RV data very well, 
there are times when the measurements deviate from this 
solution. In particular, the McD-Tull data near JD=2452600 (2002 -- 2003)
show a rapid increase in the RV, but at a later phase to the orbital solution.
Also, both the McD-Tull and TLS data at JD = 2453000 (2003 -- 2004) show considerable
scatter that is not centered on the orbital solution.

One property of long-lived, coherent periodic signals is that the L-S
power, and thus statistical significance, should increase with the number of data
points. The circles in Figure~\ref{fapgrow} show the growth of the L-S power
as a function of the number of data points taken in chronological order. 
{ The temporary decrease in power at $N$ $\approx$ 200 is related to the discrepant RV data
in the years 2002 to early 2004.}

Triangles
represent results from a simulation where 
the orbital solution was sampled in the same manner as the data and
with random noise at a level of 100 m\,s$^{-1}$ added.
The slope of the power increase for the real data is consistent with
that of a long-lived, coherent signal.

\subsection{Hipparcos photometry}

{ The Hipparcos photometry for $\alpha$~Tau 
covers the time interval from 1990.1 to 1992.6 and is therefore
contemporaneous with a part of the RV measurements.}
Figure~\ref{photphase} shows this photometry phased to the 629 d orbital
period. Although there is considerable scatter in the data there are no
obvious sinusoidal variations.  Observations of $\alpha$~Tau made with the MOST
space telescope reveal periodic  photometric variations with a peak-to-peak value of
$\approx$ 0.02 mag (Matthews, private communication) which is comparable to the
scatter in the Hipparcos photometry.

\subsection{Variations in activity indicators}

One explanation for the 629 d period  in $\alpha$ Tau is that it is due
to rotational modulation caused by surface structure. Even though this is not supported
by the previous bisector measurements, it is still worth investigating
other activity indicators. Unfortunately, only the
McD-Tull  observations 
cover the Ca II H \& K lines. For the CFHT and DAO observations we must use
the Ca II 8662 {\AA} line and
H$\alpha$ for the TLS data as activity diagnostics.

\subsubsection{Ca II 8662 {\AA} variations}
Larson et al. (1993a) showed that the Ca II $\lambda$8662 line is 
suitable for measuring chromospheric activity.  Figure~\ref{ftca}
shows the L-S periodogram of the variations in 
Ca II $\lambda$8662 equivalent width,  $\Delta$EW$_{8662}$, measured
from the CFHT and DAO data. There is no significant power at the
orbital,  or any other period.

\subsubsection{Ca II variations}

The McDonald Ca II S-index measurements from the McD-Tull data set 
show long term variations with a {possible} period
of $P$ $\approx$ 4000 d (Figure~\ref{sindex}). The FAP for 
this signal is
FAP $\approx$ 0.08 as determined
from a bootstrap analysis
(Murdoch et al. 1993; K\"urster et al. 1997). 
A sine fit to these data 
results in $P$ = 4320 $\pm$ 450 d. { This periodic signal is only
suggestive as there are insufficient data with too much scatter to be convincing.
However, the rough time scales of this possible periodic variations is
consistent with long term variations in the RV residuals (see below).}
After removing this long-term
trend the L-S periodogram shows power at a period, $P$ = 521 $\pm$ 10 d 
(top panel of Figure~\ref{indices}). 
{This peak is not significant with FAP $\approx$ 
0.2, but as we shall see it is  
close to a  value found in the residual RV and H$\alpha$ measurements.}

\begin{figure}[h]
\resizebox{\hsize}{!}{\includegraphics{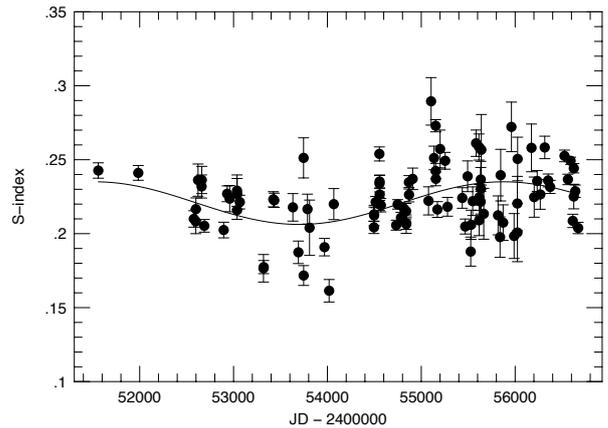}}
\caption{The variations of the McDonald S-index measurements with time.
The line represents a sine fit with a period $P$ = 4320 $\pm$ 450 d.
}
\label{sindex}
\end{figure}

\subsubsection{H$\alpha$ variations}
The TLS spectral data cover the Balmer H$\alpha$ line which is 
also useful for discerning whether RV variability is due to intrinsic
stellar variability from activity (see K\"urster et al. 2003). Thus we 
measured the equivalent width of H$\alpha$ from the TLS spectra
focusing on the time span JD = 2452656 -- 2454160 {in
order to investigate the cause for the large scatter in the residuals. }
For main sequence stars  such a measurement
is problematic because of blending of the stellar H$\alpha$ with 
telluric lines. However, in a giant
star like Aldebaran this is much easier because the H$\alpha$ line is 
so narrow. We could thus use a window of only $\pm$ 1.0 {\AA}
centered on H$\alpha$
to measure the equivalent width using a Gaussian fit to the profile. 
The two prominent H$_2$O absorption
lines at 6560.5 {\AA} and 6565.5 {\AA}, which have equivalent widths of
22 m{\AA} and 14 m{\AA}, respectively, are outside of 
the measurement window. 

\begin{figure}[h]
\resizebox{\hsize}{!}{\includegraphics{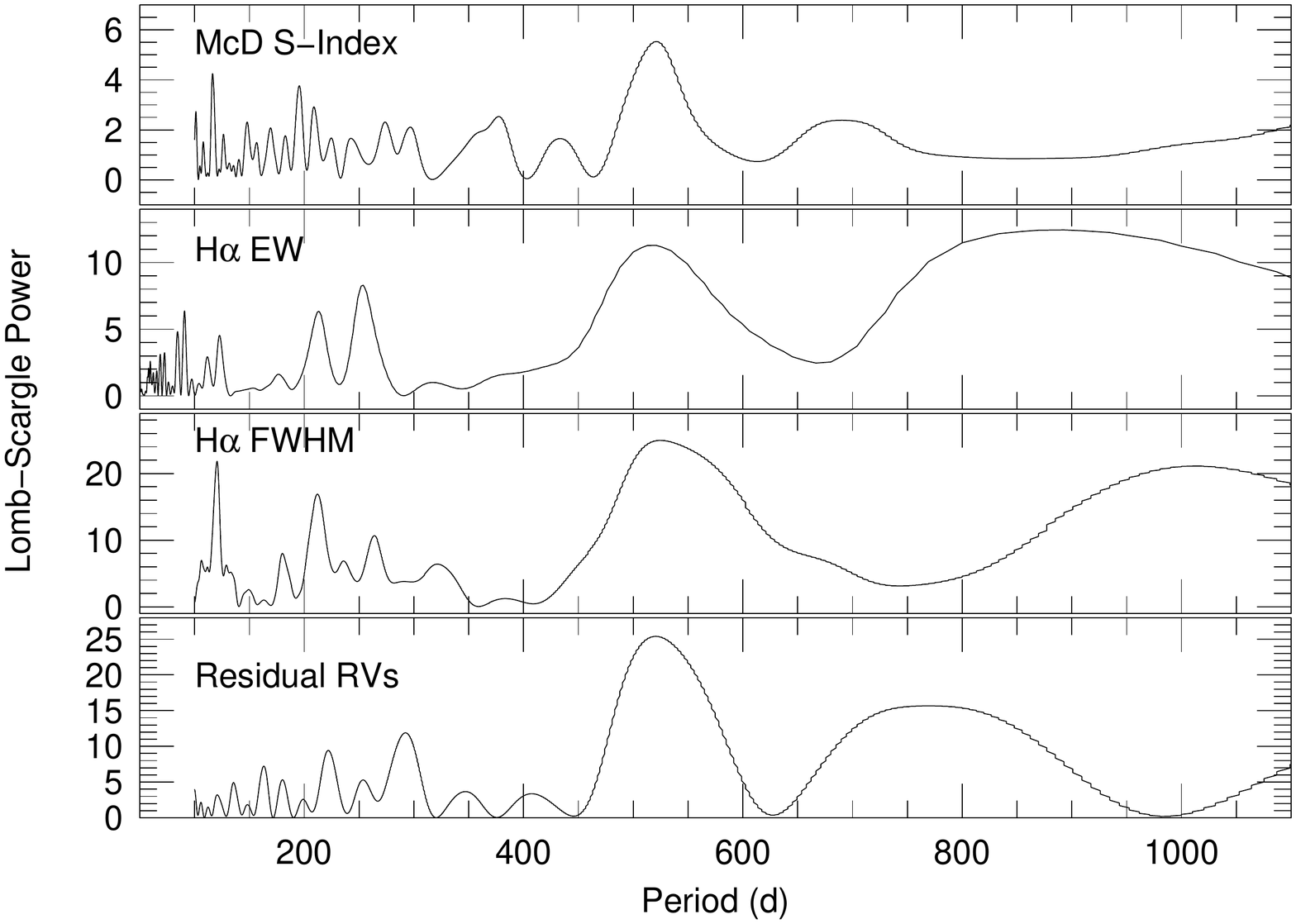}}
\caption{(top to bottom) The L-S periodograms
of the McDonald S-index, 
H$\alpha$  equivalent width, H$\alpha$ FWHM, and residual
RV   measurements. {The RV residuals were taken from the
time interval 2004.7 -- 2008.2 (see text). }}
\label{indices}
\end{figure}

The middle-upper panel of Figure~\ref{indices} shows the L-S periodogram of the H$\alpha$
equivalent width variations. There are two significant peaks, one
corresponding to a period of 892 d
and another corresponding to a period of 
520 d. These periods are clearly
aliases of each other as fitting the data with the sine function
of one period and removing it from the data causes the other to
disappear. The FAP estimated using a bootstrap of 2 $\times$ 10$^{5}$ random
shuffles of the RV data yielded
no instance when the random periodogram had more L-S than the data. The FAP
is thus less than 5 $\times$ 10$^{-5}$. 

We also measured the full width at half maximum (FWHM) of the
H$\alpha$ line. The FWHM of photospheric lines have been shown
to be good tracers of variations due to activity in active main
sequence stars (Queloz et al. 2009). The FWHM of 
H$\alpha$ may also provide a good tracer of chromospheric activity.
Furthermore, the FWHM may be less sensitive to contamination by
telluric lines as would be the case for
equivalent width measurements.  The lower-middle panel of Figure~\ref{indices} is the
L-S periodogram of the H$\alpha$ FWHM measurements. It shows
the same two dominant peaks as the equivalent width, but in this case
a peak at 526 d is stronger.

{The largest peak in the L-S periodogram of the Ca II S-index data attains
more significance when one considers that a similar period was  found in 
other activity indicators. The previous FAP of approximately 20\% was the probability
that a random data could produce a peak higher than the real data over a broad
period range. Since there is a known, and significant period in other data
we need to assess the FAP that noise produces a  higher peak {\it exactly}
at a period of interest. In this case the FAP for the peak near 520 d
in the Ca II periodogram  has a FAP 
$\approx$ 0.4\% using the prescription in Scargle (1982),
or $\approx$ 1\% using a bootstrap. The peak in the Ca II data thus seems reasonably
significant.}

\begin{figure}
\resizebox{\hsize}{!}{\includegraphics{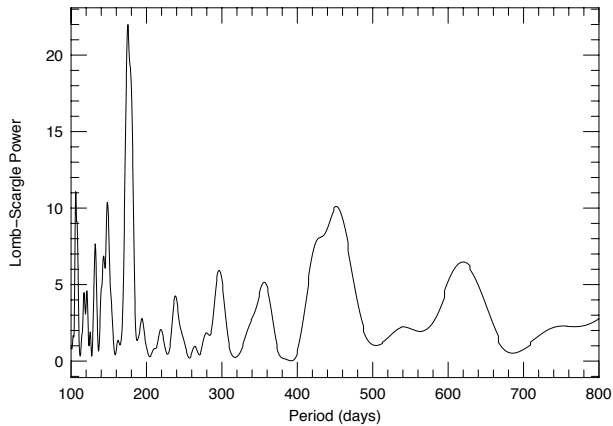}}
\caption{The L-S periodogram of the bisector span measurements.  
}
\label{bvs}
\end{figure}

\begin{figure}[h]
\resizebox{\hsize}{!}{\includegraphics{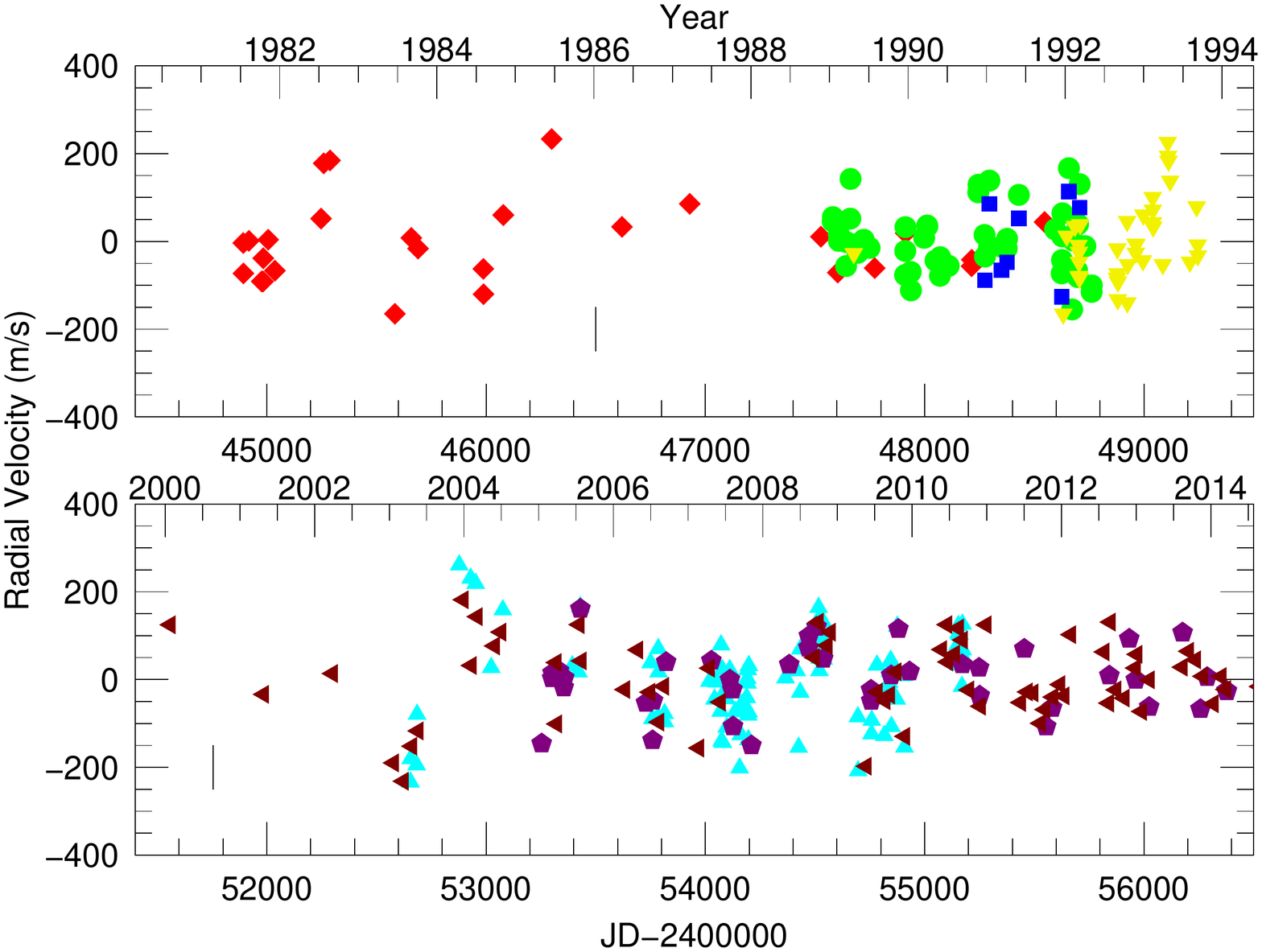}}
\caption{Residual RV variations for $\alpha$ Tau after subtracting the orbital
solution of Table 10.
}
\label{RVres}
\end{figure}

\begin{figure}
\resizebox{\hsize}{!}{\includegraphics{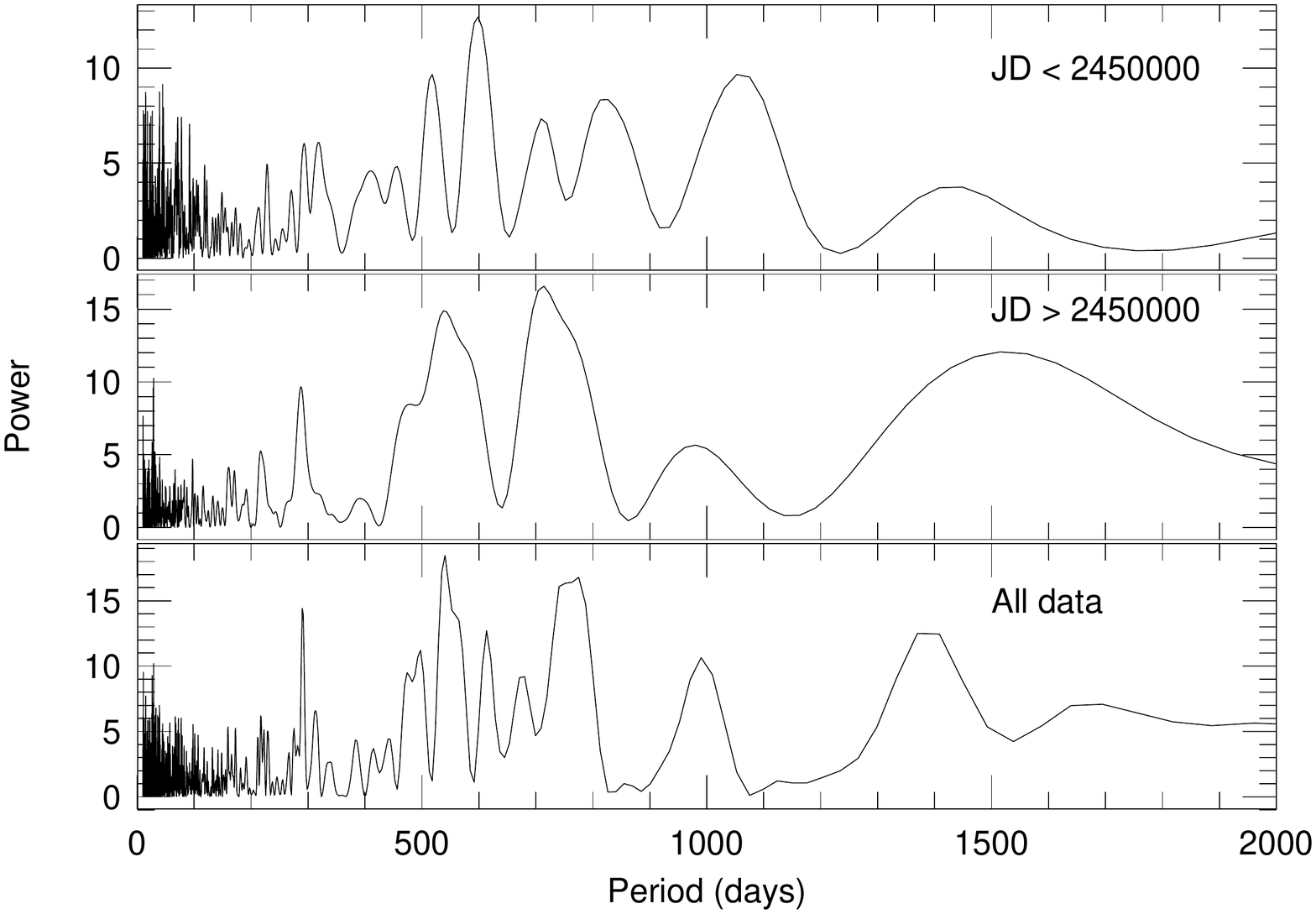}}
\caption{L-S periodograms of the residual RV data after removing the
orbital solution. The top panel is from data prior to JD = 2450000,
the middle panel for data afterwards, and the bottom panel for all data.
}
\label{RVresft}
\end{figure}

\subsection{Bisector variations}

Spectral line bisectors have become a common tool  to ensure that RV variations attributed
to orbital motion are not in fact due to intrinsic stellar variations 
(Hatzes et al. 1998; Queloz et al. 2001). We thus examined the bisectors using the
McD-Tull data set. {This was selected because we wanted
to use a consistent data set in measuring the bisectors. The McD-Tull 
data has the longest
time baseline and a pipeline has been developed for measuring bisectors from 
this instrument. Although the BOAO data has the highest resolution,
the BOES has an unstable instrument profile that is modeled in
calculating the RVs, but is not accounted for in any bisector measurements.}

In our analysis we used a statistical approach in order to select
the ``cleanest'' lines for calculating the bisector.
First,
we identified those spectral lines that were relatively
deep and that  appeared to be free from nearby blends in the
{wavelength region 4000 -- 10000 {\AA}. We ignored lines in the
spectral regions 4610 -- 4760 {\AA}, 6830 -- 6940 {\AA}, 7590 -- 7720 {\AA},
and 8130 -- 8270 {\AA}, as well as smaller regions containing telluric lines.}
A linear fit to each side of the
spectral line and a Gaussian fit to the line core was performed.
Next we examined the deviation of
the fit parameters from individual lines from that of
the average  values.
A large deviation indicated that  a
line was strongly affected by a blend and this line was removed from the analysis.
We then computed the bisector for each the final selection of 75 
spectral lines. 
The bisector velocity span was calculated  using the 
difference of the average bisector
values  between flux
levels of 0.1 -- 0.3 and 0.7 -- 0.9 of the 
continuum value. 
Finally,  the 
individual bisector span measurements were combined to produce
an  average bisector velocity span.

Figure~\ref{bvs} shows the L-S periodogram of the bisector span measurements. No significant peak is found
at the 629\,d period. However, a significant peak is found at a period, $P$ = 175.4 $\pm$ 0.6 d {
which is 1/3 the period found in the CaII S-index and H$\alpha$ measurements.}
A bootstrap analysis yields a FAP $<$ 5 $\times$ 10$^{-6}$ for this signal.

\begin{figure}
\resizebox{\hsize}{!}{\includegraphics{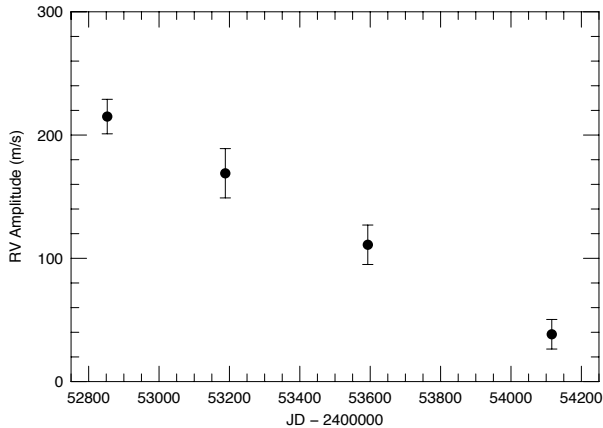}}
\caption{The K-amplitude of the 521\,d period seen in the RV residuals as a function of time.
}
\label{ampvar}
\end{figure}
 
\begin{figure}
\resizebox{\hsize}{!}{\includegraphics{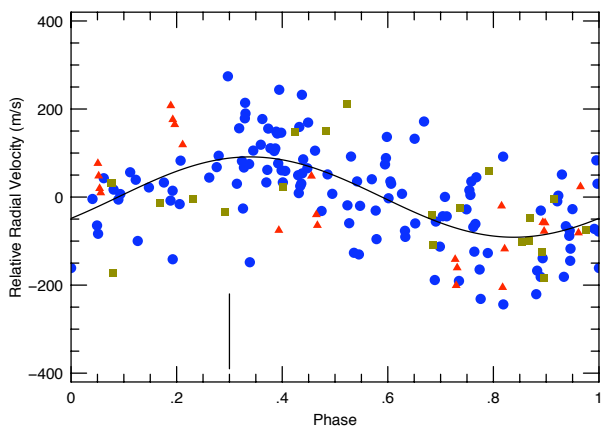}}
\caption{Residual RV variations phased using a best-fit
period of 534.7 d for three epochs:
1980.8 -- 1986.8 (triangles), 1992.6 -- 1993.6 (squares),
and 2004.7 -- 2008.2 (dots).
The vertical line indicates the peak-to-peak RV variations due to the stellar
oscillations.
}
\label{resphase}
\end{figure}

\subsection{Residual RV variations}

The activity indicators of H$\alpha$ and Ca II K show variations with 
a period of $\approx$ 520 d that are most likely caused by rotational
modulation of surface features from stellar active regions. We
suspect that such a  period in the 
RVs may be responsible for the poor
fits to the orbital solution in some years.

Figure~\ref{RVres} shows the residual RV variations after subtracting
the orbital solution. Note the sine-like variations between 2003 and 2004
(JD-2400000 = 52500 -- 53600), the same time where the orbital fit is poor. 
Figure~\ref{RVresft} shows the L-S periodograms of the
residual RV data prior to  (top) and after (middle) JD $=$ 2450000.
The lower panel shows the periodogram of all the residuals. 

The behavior of the L-S periodogram suggests that this 
period is not long-lived. Different peaks become more prominent
as more data is added, and more importantly the L-S power
does not dramatically increase in the combined data sets, unlike
the 629\,d period. The RV residuals prior to JD $=$ 2450000
show the strongest power, $z$,  at $P$ = 598 d and $P$ = 518 d ($z$ = 12.7 and
9.7, respectively).  The RV residuals after JD $=$ 2450000, on the other hand,
show the most power at $P$ = 715 d and $P$ =538 d ($z$ = 16.5 and
15.8, respectively). The combined
data show the strongest peak at $P$ = 542 d, $z$ = 18.4.  If this were a 
coherent and long-lived signal, then it should have been detected with
higher significance.   A simulated periodic signal with $P$ = 540 d and
$K$ = 100 m\,s$^{-1}$ and $\sigma$ = 100 m\,s$^{-1}$ was generated and
sampled in the same
way as the data. A periodogram analysis revealed that the signal was easily detected in just the first half
of the data ($z$ = 20) and was  highly significant in the full data set
($z$ = 60).

A visual inspection of the time series residuals also indicates that the $\sim$ 500 d
variations are not long lived. There are  strong variations at JD $\approx$ 2453000  that clearly
fade away by JD $\approx$ 2454000.  To investigate this possible amplitude variation  we
divided the residual RV data taken between 2002.8 and 2008.3 into four sliding subsets in
time with widths of $\sim$ 2 -- 2.5 yrs and found the best fit
$K$-amplitude using the  521\,d period. The $K$-amplitude decreases from
about 200 m\,s$^{-1}$ to 40 m\,s$^{-1}$ in about 3.5 years (Figure~\ref{ampvar}).

Although this signal is not long-lived, there are other instances when it appears
at other epochs.
We divided the RV residual data into three subset epochs: ``Epoch 1'', JD = 2444531 -- 2446726
(1980.8 -- 1986.8), ``Epoch 2'', JD = 2448831 -- 2450000 (1992.6 -- 1993.6), and ``Epoch 3'',
JD = 2453253 -- 2454396 (2004.7 -- 2008.2). These three epochs
showed the largest variations in the RV residuals. A period search was then performed in
each epoch. In Epoch 3 the RV residuals were best fit with $P$ = 521 $\pm$ 11 d, and 
a $K$-amplitude of 95 $\pm$ 10 m\,s$^{-1}$. In Epoch 2 the best fit was
with $P$ = 526 $\pm$ 82 d, $K$ = 132 $\pm$ 30 m\,s$^{-1}$. The data were too 
sparse in Epoch 1 for a reliable fit; however, when combining the data from all
three epochs a good fit was obtained with $P$ = 534.7 $\pm$ 1.7 d, and
$K$ = 91.2 $\pm$ 9.0 m\,s$^{-1}$. 
Figure~\ref{resphase} shows the residual RV variations of the three epochs phased to 
the 534.7\,d period. Although this signal appears intermittently in the data, it appears
to phase well with other epochs.

We should note that the peak in the L-S periodogram for the residual
RVs spanning for Epoch 3, 2004.7 -- 2008.2 (lower panel of Figure~\ref{indices}),
coincides exactly with the one found
in the Ca II S-index and H$\alpha$ measurements over the same time span.

Table 11 summarizes the  periods found by the analysis of the activity indicators
and the residual RV data. 

\section{Discussion}

Our new radial velocity measurements for $\alpha$~Tau   when combined with { previous
published}
data show that the  645\,d RV variations found by HC93 
are indeed long-lived and coherent. 
A careful examination of the
Ca II K emission, spectral line shapes and Hipparcos photometry
reveals no convincing variation with the 629\,d RV period. A previous study by Hatzes \& 
Cochran (1998) established that there were no spectral line shape variations (line bisectors)
with the 629\,d RV period. All of these support the fact that the
RV variations at this period are due to the planet hypothesis first proposed by HC93.

The analysis of the ancillary data of residual RVs, Ca II K S-index, and H$\alpha$ reveals the presence
of a $\approx$ 520\,d  period that seemed  to be only in the residual RV data taken
between 2004.7 and 2008.2, the early McDonald and DAO data (1992.6--1993.6), and possibly 
the CFHT data (1980.8 -- 1986.8).
This period is not present in the
RV data or Ca II 8662 {\AA} measurements made prior to 2000. 
We interpret this signal as due
to rotational modulation by stellar surface structure. 
The amplitude variations (Fig.~\ref{ampvar})
can be explained by the evolution of surface spot features. A reasonable explanation
for the absence of the signal in some years is an activity cycle.
If indeed the $\sim$ 520\,d period is due to rotation then this provides further
support to the 629\,d RV period arising from Keplerian motion from a companion.

One could argue that both the 629\,d period seen in the RV data and the 520\,d period seen in the
residual RVs and activity indicators are all due to rotation and we are seeing the effect of differential rotation.
However, if the 629\,d period is also due to rotational modulation, then we should have at least some
evidence of its presence in the activity indicators  or 
the line bisector measurements  of 
Hatzes \& Cochran (1998) as is the case for the 520\,d period. Stellar surface structure that cause RV variations should also produce variations
in at least one of these other quantities. 

If indeed the two periods are due to differential rotation then we can estimate the  differential 
rotation parameter, $\alpha$ = ($\omega_{pole}$ $-$ $\omega_{equator}$)/$\omega_{equator}$ where $\omega_{pole}$ and 
$\omega_{equator}$ are the polar and equatorial rotational angular velocities.  Taking our two periods
as the polar and equatorial rotational periods results in $\alpha$ $\ge$ 0.2, or greater
than  the
solar value of $\alpha$  = 0.14 . The value of
$\alpha$ could be considerably larger if the one of the periods represents a mid-latitude rotational
period rather than the polar one. This large  value of differential rotation should result in
strong stellar activity, but our Ca II and H$\alpha$ measurements suggest a relatively low level
of activity for this star. 

However,  recent work by K\"uker \& R\"udiger (2012) on the K giant star KIC 8366239 predicted a 
solar type differential rotation in this star, so modest amount of differential rotation in
K giant stars may be possible. Looking at the RV variations between 2003 and later, 
if one were to explain the long-period RV variations of 
$\alpha$ Tau, then you require a  spot feature to form in 2003 at a latitude  where the rotational
period is $\sim$ 500 d. This spot would quickly ($\Delta$$t$ $\approx$ 1 yr) have to migrate
to another latitude that has rotational period of 630 d. It would then have to be stable in size
and location for  the next eight years. 
Given our limited knowledge of activity in K giant stars this may be possible, but seems less plausible
than having a sub-stellar companion and rotational modulation.

The spectral line bisector variations show no significant variations at either the orbital period
of the planet or the $\approx$ 520 d variations that we attribute to the rotation of the star.
However, there is a significant peak at a period of 175 d. Interestingly, this is one-third
the period found in the RV residuals and the activity indicators (Table 11). The most logical 
explanation for this period is that it is the third harmonic of the stellar rotation period, i.e.
$P_{rot}$/3.

It is not clear why we see variations in the bisectors at third harmonic of the rotational period.
Naively, one expects that the variations in spectral line shapes (bisectors) are directly
tied to the RV variations and that one should at least see evidence of $P_{rot}$ in the
bisectors. However, we should stress that we know little about the nature of the activity
related structure (e.g. starspots) and the velocity field (e.g. convection) on the
surface of giant stars and how these influence both the line shapes and the integrated
RV measurements. It may be that a harmonic of the rotational period is only seen in one 
quantity. Alternatively, the lack of power at $P_{rot}$ in the periodogram of the bisector
variations may be due to a combination  of the real variations, the sampling, and the noise
characteristics of the bisector measurements.

There is evidence, however, of a harmonic of the rotational period in a K giant star being
present in some observed quantities. In a pioneering study of the He I 10830 {\AA} line in 
Arcturus,  Lambert (1987) found evidence for a 78 d variation in this activity
indicator. Lambert argued that the He I variations represent one-third of the actual
stellar rotational period of 233 d. Subsequently, HC93 found evidence of an RV period
of 231 d in Arcturus, nearly identical to the inferred rotational period.
Unlike $\alpha$ Tau and $\beta$ Gem, the long period variations in Arcturus found by HC93 
are most likely due to rotational modulation.

Spectral line bisector measurements have become a standard tool for confirming planets
around giant stars. The assumption is that if 
one sees no variations in the bisectors, or variations at a different period to the RV period, then the
RV variations must be due to a planet. In the case of $\alpha$ Tau, one sees periodic bisector 
variations different to the 520\,d residual RV  period, but at a harmonic. The measurements
of activity indicators show, however, that the residual RV and bisector periods {
can be interpreted in a consistent way if one assumes that they are
all related to the rotational period of the star.}
There are other instances where
harmonics of the rotational period of the star have masqueraded as an RV signal believed to
be due to a planet (Robertson et al. 2014).  In confirming planets around giant stars
it is wise to examine all available activity indicators.

The residual RVs show indications for an activity cycle in $\alpha$ Tau. These show an active phase
that lasts approximately 3.5 years (Fig.~\ref{ampvar}) after which the star enters a quiet phase.
The activity-related RV variations were strongest in 2003 -- 2005, 1993, and possibly  
around 1984 judging by the large
residual variations in the CFHT data. This implies an activity cycle of $\approx$ 10 years. 
However, given the sometimes large gaps in our  RV coverage, this time scale {is only a crude estimate.}
Interestingly, this is near the period seen in the long-term variations in the S-index
(Fig.~\ref{sindex}).
The decay in the RV amplitude shown in Figure~\ref{resphase} indicates a ``spot'' lifetime
of $\approx$ 3.5 years. 

It is not known what the nature of the activity is on a K giant, whether these are cool spots, or possibly 
large convection cells organized by magnetic fields. The variations seen in the traditional indicators
such as H$\alpha$ suggest this activity is analogous to solar magnetic activity. The fact
that subsets of the residual RV data separated by about ten years 
phase well when using the same period
(Fig.~\ref{resphase}) suggests that this activity may occur in active longitudes on the star. 
Dedicated RV monitoring of K giant stars with better temporal sampling may provide valuable clues
to stellar activity on evolved stars.  

All the available evidence suggests that the 629\,d RV variations
in $\alpha$ Tau are most likely caused by Keplerian motion of a companion. The additional 
$\sim$ 520 d period that is short lived and not coherent is due to rotational modulation 
of active regions. 

Our value  of 1.13 $M_\odot$ for the stellar mass results in a minimum companion mass of
6.47 $\pm$ 0.53 $M_{Jup}$. The substellar companion to $\alpha$ Tau has similar properties to other 
companions around giant stars. These are mostly 
``super planets" in the mass range 3 -- 14 M$_{Jup}$ and with orbital radii of 
$\approx$ 2 AU (Hatzes 2008).  

Our RV and ancillary measurements for $\alpha$ Tau
demonstrate that K giant stars can have sub-stellar companions {\it and} stellar surface structure, both of
which show up as long (several hundreds of days) periods. This presents a cautionary tale
for  programs searching for planets around evolved stars with the RV method. Our residual
RV measurements show evidence for variations with a 520 -- 540 d period and a $K$-amplitude of
$\approx$ 100 m\,s$^{-1}$. This is comparable to the  amplitudes and periods found in the RV variations of
other K giant stars. If you had poor temporal sampling and were to phase the residual RV from approximately
a decade earlier,  the results would suggest a long-lived coherent
RV signal consistent with a planetary companion having a minimum mass of approximately 8 $M_{Jup}$. 
The L-S periodogram of the data when this signal was present yields power 
at this period consistent with
a FAP $\approx$ 10$^{-10}$. A bootstrap analysis yields a FAP $<$ 5 $\times$ 10$^{-6}$.
However, this ``planetary-like'' signal
disappears with more data having better  sampling. 

One should be cautious
in interpreting RV variations as due to companions in giant stars. This is especially true
when one does not have the luxury of three decades of data to establish longevity of the
RV signal, a primary requirement for a planetary signal. Looking at ancillary
measurements of activity indicators (H$\alpha$, Ca II, photometry, line shapes, etc.)
is  necessary to confirm signals from companions. In particular, it appears
that H$\alpha$ is a suitable diagnostic of activity variations. Clearly, K giant stars 
represent a fertile ground not only for discovering exoplanets, but also for investigating stellar activity in evolved stars.

\begin{acknowledgements}
The authors wish to thank the referee, Martin K{\"urster}, for his very careful
reading of the manuscript and for making suggestions that greatly improved
the paper. Referees like him are greatly appreciated.
APH and MD acknowledge
grant HA 3279/8-1 from the Deutsche Forschungsgemeinschaft
(DFG). WDC acknowledges the support of NASA grants NNG04G141G and NNG05G107G.
WDC, ME, and PM acknowledge support from grant AST-1313075 from the 
National Science Foundation.
GAHW is supported by the Natural Sciences and Engineering Research Council 
of Canada.  AES gratefully acknowledges partial funding from the 
Sciex Programme of the Rector's Conference of
the Swiss Universities.

We also thank the many observers of the McDonald Observatory planet 
search program that have helped to obtain the extensive $\alpha$~Tau 
data: Paul Robertson, Erik Brugamyer, Rob Wittenmyer, Kevin Gullikson, 
Ivan Ramirez, Marshall Johnson, Caroline Caldwell, Diane Paulson and 
Candace Gray.

This research has made use of the SIMBAD data base operated
at CDS, Strasbourg, France.

\end{acknowledgements}

\clearpage

\begin{table}[h]
\begin{center}
\begin{tabular}{ll}
Parameter & Value  \\
\hline
$\mathrm{Spectral\,\,type}$	        & $\mathrm{K5III}$     	\\
$m_{V}$ [mag]   			& 0.85    		\\
$M_{V}$ [mag]     		        & $-$0.65 $\pm$ 0.041	\\
$B-V$ [mag]			        & 1.520 $\pm$ 0.005	\\
$\mathrm{Parallax}$ [mas]		& 48.94 $\pm$ 0.77 	\\
$\mathrm{Distance}$ [pcs]		& 20.43 $\pm$ 0.32  	\\
$\mathrm{Mass}$ [$\mathrm{M}_{\sun}$]	& 1.13  $\pm$ 0.11 	\\
$R_{*}$ [$\mathrm{R}_{\sun}$]           & 45.1  $\pm$ 0.1       \\
$\mathrm{Age}$ [$\mathrm{Gyr}$]	        & 6.6   $\pm$ 2.4 	\\
$T_{\mathrm{eff}}$ [$\mathrm{K}$] 	& 4055  $\pm$ 70 	\\
$\mathrm{[Fe/H]}$ [$\mathrm{dex}$]	& $-$0.27 $\pm$ 0.05    \\
$\log{g}$ [$\mathrm{dex}$]	        & 1.2  $\pm$ 0.1        \\
\hline
\end{tabular}
\caption{Stellar parameters of $\alpha$ Tau.}
\end{center}
\label{parameters}
\end{table}

\begin{table}
\begin{center}
\begin{tabular}{ccccc}
Data Set  &  Coverage & Tech. & N & O$-$C    \\
          &  (Year)   &           &   & (m\,s$^{-1}$) \\
\hline
CFHT                 & 1980.80--1991.59 & HF      & 29      & 90   \\
DAO                  & 1989.02--1993.65 & HF      & 36      & 91 \\
McD-2.1m             & 1988.73--1992.25 & I$_2$   & 50      & 72     \\
McD-CS11             & 1990.78--1992.06 & I$_2$   & 9       & 92    \\
McD-Tull             & 2000.01--2013.92 & I$_2$   & 104     & 86   \\
TLS                  & 2003.04--2010.83 & I$_2$   & 103     & 98    \\
BOAO                 & 2004.68--2013.71 & I$_2$   & 42      & 72   \\
\hline
\end{tabular}
\caption{The data sets used in the orbital solution.}
\end{center}
\label{data}
\end{table}

\begin{table}
\begin{center}
\begin{tabular}{lrr}
Julian Day   & RV (m\,s$^{-1}$)  & $\sigma$ (m\,s$^{-1}$) \\
\hline
2444531.0741  & 131.0  & 20.8 \\
2444532.1116  & 61.2   & 13.5 \\
2444559.0096  & 115.1  & 15.6 \\
2444621.8398  & -43.0  & 14.8 \\
2444628.7987  & -51.3  & 2.5 \\
2444629.7511  & 0.5    & 3.4 \\
2444653.9218  & 14.2   &  9.9 \\
2444686.7929  & -94.4  & 22.3 \\
2444914.0599  & -44.4  & 37.5 \\
2444926.9848  & 96.7   & 35.2 \\
2444958.0170  & 148.3  & 21.2 \\
2445277.0688  & -147.6 & 24.9 \\
2445357.7746  & -64.4  & 11.1 \\
2445390.8439  & -118.1 & 4.12 \\
2445711.8903  & 76.5   & 16.2 \\
2445712.9192  & 19.4   & 23.7 \\
2445809.7134  & 180.2  & 15.4 \\
2445810.7059  & 179.0  & 17.6 \\
2446048.0380  & 112.5  & 9.2 \\
2446393.9708  & 178.9  & 25.5 \\
2446726.9944  & -49.9  & 11.4 \\
2447372.0968  & -123.1 & 8.7 \\
2447455.0482  & -135.7 & 16.3 \\
2447545.8320  & 71.7   & 6.3 \\
2447636.7040  & 87.9   & 11.6 \\
2447789.0919  & 40.7   & 6.5 \\
2448113.1188  & -75.6  & 10.1 \\
2448114.1145  & -59.2  & 7.1 \\
2448472.1388  & 3.6    & 8.1 \\
\hline
\end{tabular}
\end{center}
\caption{CFHT RV measurements for $\alpha$ Tau}
\label{cfht}
\end{table}

\begin{table}
\begin{center}
\begin{tabular}{ccc}
Julian Day   & RV (m\,s$^{-1}$)  & $\sigma$ (m\,s$^{-1}$) \\
\hline
2447533.6953  &  39.3   & 32.3 \\
2448563.8785  & -284.2  & 22.0 \\
2448580.8506  & -114.8  & 59.6 \\
2448621.8210  & -98.6   & 14.8 \\
2448638.7093  & -138.1  & 12.4 \\
2448639.6839  & -208.6  & 32.5 \\
2448640.7434  & -153.1  & 13.1 \\
2448641.7160  & -175.8  & 11.8 \\
2448643.6215  & -211.2  & 25.4 \\
2448647.7046  & -87.5   & 16.9 \\
2448828.9868  & 34.3    & 21.9 \\
2448829.9920  & 101.8   & 33.3 \\
2448831.0140  & 46.3    & 14.8 \\
2448831.9793  & -11.9   & 19.3 \\
2448833.0160  & 29.5    & 14.0 \\
2448877.9797  & 195.0   & 38.7 \\
2448879.0493  & 9.9     & 30.7 \\
2448881.0524  & 97.3    & 22.7 \\
2448919.7802  & 137.8   & 25.3 \\
2448921.0926  & 116.7   & 24.3 \\
2448922.0433  & 135.0   & 19.9 \\
2448955.8471  & 76.8    & 24.8 \\
2448957.8805  & 179.2   & 30.8 \\
2449003.7779  & 173.8   & 15.0 \\
2449004.5585  & 144.3   & 18.2 \\
2449005.5411  & 114.5   & 16.3 \\
2449006.6397  & 103.3   & 19.8 \\
2449053.5878  & -36.7   & 22.0 \\
2449077.6181  & 213.4   & 15.2 \\
2449079.6309  & 179.6   & 18.0 \\
2449081.6194  & 166.0   & 11.5 \\
2449089.6298  & 111.5   & 35.2 \\
2449187.0175  & -162.5  & 24.0 \\
2449220.0447  & -51.3   & 15.8 \\
2449225.0493  & -139.7  & 30.8 \\
2449226.0483  & -164.3  & 24.3 \\
\hline \\
\end{tabular}
\end{center}
\caption{DAO RV measurements for  $\alpha$ Tau}
\label{dao}
\end{table}

\begin{table}
\begin{center}
\begin{tabular}{ccc}
Julian Day   & RV (m\,s$^{-1}$)  & $\sigma$ (m\,s$^{-1}$) \\
\hline
2447429.9795 &  -57.2   & 11.0  \\
2447430.8745 &  -46.3   & 7.5  \\
2447459.9282 &  -46.8   & 18.6  \\
2447460.7788 &  -61.6   & 20.0  \\
2447495.8505 &  -61.7   & 15.0  \\
2447496.8154 &  -5.3    & 15.0  \\
2447516.7890 &  82.5    & 15.0  \\
2447517.8105 &  174.7   & 12.0  \\
2447551.7055 &  79.8    & 12.0  \\
2447552.6005 &  60.4    & 15.0  \\
2447582.6601 &  124.7   & 15.0  \\
2447611.6093 &  124.2   & 12.0  \\
2447785.9804 &  -57.4   & 12.0  \\
2447786.9423 &  -3.8    & 5.0  \\
2447787.8564 &  49.4    & 8.8  \\
2447813.9350 &  -83.3   & 12.0  \\
2447814.8994 &  -126.5  & 10.0  \\
2447879.8295 &  -76.6   & 12.0  \\
2447894.7841 &  -62.1   & 15.0  \\
2447895.7368 &  -64.6   & 12.0  \\
2447935.6987 &  -171.2  & 12.0  \\
2447957.5781 &  -215.4  & 15.0  \\
2447958.6899 &  -173.5  & 9.0  \\
2447959.6293 &  -194    & 7.5  \\
2448000.6250 &  -195.9  & 7.5  \\
2448145.9926 &  142.9   & 12.0  \\
2448146.9980 &  162.0   & 12.0  \\
2448176.9609 &  95.0    & 20.0  \\
2448178.9414 &  48.5    & 12.0  \\
2448198.9951 &  97.1    & 12.0  \\
2448200.9843 &  249.1   & 15.0  \\
2448260.7382 &  130.1   & 10.0  \\
2448285.7705 &  121.6   & 12.0  \\
2448287.7919 &  142.9   & 12.0  \\
2448345.6191 &  201.7   & 12.0  \\
2448523.8994 &  -71.0   & 7.5  \\
2448555.9746 &  -195.3  & 15.0  \\
2448557.8896 &  -166.1  & 7.6  \\
2448558.7910 &  -112.5  & 14.6  \\
2448559.8574 &  -60.7   & 5.1  \\
%
2448591.6933 &  28.0    & 7.4  \\
2448607.7333 &  -296.5  & 7.7  \\
2448613.8564 &  -140.2  & 12.0  \\
2448631.8134 &  -205.7  & 7.5  \\
2448635.7939 &  -219.8  & 5.0  \\
2448636.8105 &  -99.0   & 5.0  \\
2448644.7685 &  -4.9    & 7.5  \\
2448672.7177 &  -128.3  & 7.5  \\
2448703.6240 &  -198.9  & 7.6  \\
2448704.6035 &  -182.7  & 15.0  \\  
\hline \\
\end{tabular}
\caption{McDonald 2.1m telescope measurements (McD-2.1m) for $\alpha$ Tau.}
\end{center}
\label{82in}
\end{table}

\begin{table}
\begin{center}
\begin{tabular}{lrr}
Julian Day   & RV (m\,s$^{-1}$)  & $\sigma$ (m\,s$^{-1}$) \\
\hline
2448178.941 &   7.1   & 7.0 \\
2448200.989 & 208.1   & 7.0 \\
2448260.735 & 89.1    & 3.8 \\
2448287.795 & 101.9   & 6.0 \\
2448345.621 & 160.7   & 8.6 \\
2448555.973 & -236.2  & 4.8 \\
2448591.675 & -12.9   & 7.4 \\
2448607.730 & -337.5  & 4.2 \\
2448644.770 & -45.9   & 8.9 \\
\hline
\end{tabular}
\caption{McDonald 2.7m telescope measurements (McD-CS11) for $\alpha$ Tau.}
\end{center}
\label{107in}
\end{table}

\begin{table}
\begin{center}
\begin{tabular}{lrr}
Julian Day   & RV (m\,s$^{-1}$)  & $\sigma$ (m\,s$^{-1}$) \\
\hline
2451558.7500 & 154.2  & 2.8 \\
2451984.5664 & 95.5   & 6.4 \\
2452576.9765 & -104.4 & 3.7 \\
2452597.8789 & -43.6  & 5.0  \\
2452598.8984 & -108.8 & 5.5 \\
2452621.8632 & -95.4  & 6.5 \\
2452659.7343 & -21.4  & 5.4 \\
2452661.7148 & 20.6   & 5.4 \\
2452688.7148 & 58.3   & 5.3 \\
2452689.6015 & 1.0    & 4.5 \\
2452894.9765 & 122.4  & 3.8 \\
2452931.9179 & -61.7  & 4.6 \\
2452958.8281 & 29.9   & 3.7 \\
2453036.7304 & -79.0  & 5.7 \\
2453037.5703 & -64.4  & 3.8 \\
2453038.5781 & -16.0  & 5.6 \\
2453066.6250 & -7.2   & 4.1 \\
2453318.8789 & 185.1  & 5.6 \\
2453320.8320 & 44.5   & 3.9 \\
2453423.5781 & 180.3  & 3.9 \\
2453433.5781 & 86.1   & 4.5 \\
2453631.9453 & -154.3 & 6.8 \\
2453690.8945 & -51.8  & 5.8 \\
2453745.6992 & -95.8  & 3.2 \\
2453746.7226 & -81.5  & 5.2 \\
2453787.5742 & -89.8  & 5.7 \\
2453809.6328 & 29.8   & 5.5 \\
2453968.9570 & -20.6  & 6.6 \\
2454017.9687 & 119.1  & 6.5 \\
2454067.8437 & -13.9  & 5.4 \\
2454495.5781 & 215.6  & 3.0 \\
2454496.5703 & 109.0  & 4.0 \\
2454497.5859 & 203.2  & 4.0 \\
2454514.7031 & 270.8  & 3.3 \\
2454553.6445 & 318    & 3.7 \\
2454554.6054 & 183.4  & 2.5 \\
2454555.6132 & 188.8  & 3.1 \\
2454556.6054 & 211.5  & 3.0  \\
2454557.5937 & 243.0  & 3.8 \\
2454569.6054 & 256.6  & 3.1 \\
\hline \\
\end{tabular}
\caption{McDonald 2.7m telescope + Tull Spectrograph  measurements (McD-Tull) for $\alpha$ Tau.}
\end{center}
\end{table}

\begin{table}
\begin{center}
\setcounter{table}{6}
\begin{tabular}{lrr}
Julian Day   & RV (m\,s$^{-1}$)  & $\sigma$ (m\,s$^{-1}$) \\
\hline
2454732.9492 & -202.6 & 4.8 \\
2454783.8789 & -89.73 & 3.0 \\
2454816.8203 & -129.8 & 2.6 \\
2454820.6718 & -154.0 & 6.5 \\
2454838.5781 & -149.8 & 3.8 \\
2454839.7265 & -183.5 & 3.9 \\
2454840.7500 & -100.1 & 4.0 \\
2454868.7304 & -155.4 & 3.0 \\
2454869.7187 & -60.1  & 3.2 \\
2454908.6132 & -261.8 & 3.5 \\
2455076.9179 & 129.0  & 3.3 \\
2455100.9570 & 222.0  & 4.7 \\
2455104.8750 & 142.2  & 3.7 \\
2455135.9570 & 192.8  & 5.7 \\
2455152.8750 & 240.3  & 3.9 \\
2455153.8906 & 300.4  & 3.0 \\
2455154.7382 & 248.5  & 3.6 \\
2455171.7500 & 241.0  & 3.3 \\
2455200.7812 & 124.5  & 3.3 \\
2455254.5937 & 52.9   & 3.3 \\
2455279.6054 & 214.0  & 3.0 \\
2455436.9960 & -136.8 & 4.7 \\
2455467.9492 & -137.1 & 4.0 \\
2455493.9218 & -154.0 & 2.5 \\
2455525.8320 & -311.8 & 3.5 \\
2455526.8437 & -308.6 & 2.6 \\
2455529.7695 & -75.93 & 3.1 \\
2455547.7968 & -199.5 & 5.0 \\
2455584.7187 & -153.4 & 3.1 \\
2455614.6562 & -80.2  & 3.0 \\
2455616.5820 & -93.1  & 3.1 \\
2455617.5820 & -107.6 & 3.5 \\
2455632.6367 & -78.1  & 3.2 \\
2455633.6250 & -91.2  & 3.1 \\
2455634.5859 & -98.4  & 4.2 \\
2455639.5976 & -108.3 & 3.4 \\
2455665.5937 & 94.3   & 3.2 \\
2455817.9218 & 213.8  & 4.3 \\
2455838.9531 & 90.9   & 4.4 \\
2455845.9414 & 272.1  & 3.2 \\
\hline \\
\end{tabular}
\caption{McD-Tull measurements  for $\alpha$ Tau (cont.).}
\end{center}
\end{table}

\begin{table}
\setcounter{table}{6}
\begin{center}
\begin{tabular}{lrr}
Julian Day   & RV (m\,s$^{-1}$)  & $\sigma$ (m\,s$^{-1}$) \\
\hline
2455871.9492 & 100.0  & 3.2 \\
2455909.7539 & 45.9   & 5.7 \\
2455960.6953 & 56.6   & 3.2 \\
2455967.6250 & 79.7   & 3.1 \\
2455988.6601 & -75.2  & 3.2 \\
2456021.5937 & 68.2   & 3.0 \\
2456023.5898 & -58.5  & 3.1 \\
2456024.5859 & -90.0  & 3.9 \\
2456026.5820 & -92.7  & 3.2 \\
2456173.9570 & -103.4 & 3.5 \\
2456202.8906 & -56.3  & 4.1 \\
2456234.8671 & -49.3  & 2.5 \\
2456267.6679 & -44.1  & 3.4 \\
2456315.6992 & -26.5  & 3.7 \\
2456351.6406 & 93.6   & 3.0 \\
2456372.6406 & 91.6   & 3.1 \\
2456525.9843 & 85.2   & 3.2 \\
2456562.0039 & 29.2   & 3.5 \\
2456588.8906 & -36.3  & 5.4 \\
2456593.0156 & -65.1  & 3.1 \\
2456614.7929 & -68.3  & 3.2 \\
2456624.7617 & 114.1  & 5.7 \\
2456625.7773 & 32.1   & 3.1 \\
2456640.7070 & -26.9  & 3.0 \\
\end{tabular}
\end{center}
\caption{McD-Tull measurements  for $\alpha$ Tau (cont.).}
\label{cs23}
\end{table}

\begin{table}
\begin{center}
\begin{tabular}{lrr}
JD   & RV (m/s)  & $\sigma$ (m/s) \\
\hline
2452655.1992   & -60.9   & 3.8 \\
2452656.2109   & -7.7    & 4.1 \\
2452683.2031   & -32.0   & 3.1 \\
2452685.2421   & 81.8    & 2.9 \\
2452877.5468   & 198.7   & 12.5 \\
2452929.6054   & 122.1   & 5.2 \\
2452952.5312   & 95.3    & 2.9 \\
2453023.3125   & -115.8  & 2.9 \\
2453076.2500   & 38.3    & 4.0 \\
2453393.2226   & 106.9   & 2.6 \\
2453421.2929   & 56.3    & 2.3 \\
2453429.3515   & 194.5   & 5.1 \\
2453750.4570   & -28.7   & 2.7 \\
2453758.3242   & -143.1  & 2.2 \\
2453783.2421   & -93.4   & 2.8 \\
2453785.3164   & 66.5    & 3.1 \\
2453786.4218   & 15.5    & 3.0 \\
2453814.2656   & -21.7   & 2.8 \\
2453815.2656   & -38.6   & 3.0 \\
2454018.4765   & 76.8    & 7.8 \\
2454041.6289   & 6.1     & 5.9 \\
2454047.6132   & 81.3    & 4.9 \\
2454052.4179   & 31.1    & 6.4 \\
2454069.5625   & -57.5   & 3.1 \\
2454071.4375   & 91.4    & 2.4 \\
2454074.3554   & -130.9  & 2.4 \\
2454077.5429   & 27.2    & 1.3 \\
2454079.3867   & -140.7  & 2.3 \\
2454080.5703   & -31.9   & 3.1 \\
2454096.4062   & -61.9   & 1.3 \\
2454097.2773   & -127.8  & 1.5 \\
2454109.3515   & -48.9   & 2.9 \\
2454111.2968   & -12.8   & 5.5 \\
2454126.3632   & -155.4  & 1.7 \\
2454136.4414   & -135.4  & 4.4 \\
2454156.2890   & -283.3  & 3.2 \\
2454157.2578   & -151.4  & 2.7 \\
2454161.4179   & -212.2  & 4.4 \\
2454164.3515   & -145.8  & 3.7 \\
2454168.3632   & -149.4  & 1.7 \\
\hline \\
\end{tabular}
\caption{TLS  measurements for $\alpha$ Tau.}
\end{center}
\end{table}

\begin{table}
\setcounter{table}{7}
\begin{center}
\begin{tabular}{lrr}
JD   & RV (m/s)  & $\sigma$ (m/s) \\
\hline
2454188.3281   & -149.5  & 4.9 \\
2454191.2734   & -108.3  & 4.4 \\
2454192.2656   & -119.7  & 4.4 \\
2454193.2695   & -89.8   & 4.2 \\
2454195.2812   & -194.6  & 4.2 \\
2454196.2773   & -251.3  & 4.3 \\
2454197.3046   & -189.0  & 5.9 \\
2454198.2890   & -83.7   & 4.1 \\
2454365.6015   & -83.5   & 23.4 \\
2454425.4453   & -136.3  & 13.0 \\
2454432.4609   & 3.6     & 7.9 \\
2454433.4531   & 104.3   & 13.1 \\
2454516.2656   & 329.3   & 10.0 \\
2454520.2851   & 187.6   & 10.2 \\
2454521.2421   & 216.5   & 8.3 \\
2454522.2421   & 310.8   & 10.9 \\
2454524.3750   & 267.2   & 8.8 \\
2454528.2812   & 262.4   & 10.5 \\
2454529.3125   & 268.6   & 8.9 \\
2454530.2539   & 307.0   & 12.1 \\
2454531.2656   & 302.8   & 8.9 \\
2454532.2929   & 235.2   & 9.5 \\
2454535.2656   & 305.5   & 12.6 \\
2454539.3437   & 219.0   & 12.3 \\
2454695.6171   & -65.9   & 13.2 \\
2454696.5898   & -190.1  & 10.9 \\
2454756.6875   & -94.5   & 21.4 \\
2454757.6953   & -177.8  & 15.5 \\
2454758.7031   & -148.1  & 15.6 \\
2454759.7031   & -92.9   & 16.2 \\
2454781.6250   & -110.2  & 10.3 \\
2454782.5625   & -46.1   & 12.3 \\
2454815.5703   & -234.5  & 3.2 \\
2454819.3945   & -151.5  & 3.4 \\
2454837.4101   & -128.3  & 1.3 \\
2454840.3398   & -143.3  & 1.9 \\
2454841.2304   & -83.8   & 1.9 \\
2454842.3085   & -106.2  & 1.9 \\
2454843.3437   & -147.1  & 1.9 \\
2454844.2421   & -152.7  & 2.0 \\
\hline \\
\end{tabular}
\caption{TLS  measurements for $\alpha$ Tau (cont.).}
\end{center}
\end{table}

\begin{table}
\setcounter{table}{7}
\begin{center}
\begin{tabular}{lrr}
JD   & RV (m/s)  & $\sigma$ (m/s) \\
\hline
2454845.3164   & -82.7   & 1.3 \\
2454847.3710   & -154.6  & 1.4 \\
2454848.2929   & -234.9  & 1.9 \\
2454872.3007   & -182.8  & 18.2 \\
2454875.2421   & -17.9   & 2.3 \\
2454904.3242   & -134.9  & 2.2 \\
2454908.3359   & -297.3  & 1.9 \\
2455150.6250   & 247.4   & 6.3 \\
2455151.6171   & 241.0   & 7.4 \\
2455153.5585   & 292.3   & 4.7 \\
2455154.5859   & 275.6   & 9.7 \\
2455155.4921   & 203.5   & 3.0 \\
2455156.5507   & 229.1   & 1.9 \\
2455158.5000   & 273.8   & 3.9 \\
2455161.5195   & 226.7   & 3.3 \\
2455162.4570   & 210.2   & 2.5 \\
2455163.4765   & 223.3   & 2.6 \\
2455164.6640   & 228.5   & 8.8 \\
2455165.6171   & 227.2   & 4.0 \\
2455170.4492   & 157.2   & 1.5 \\
2455172.5664   & 298.8   & 4.5 \\
2455173.3906   & 240.2   & 2.5 \\
2455175.3984   & 223.7   & 3.0 \\
\hline
\end{tabular}
\caption{TLS RV measurements (nightly averages) for $\alpha$ Tau (cont.)}
\end{center}
\label{tls}
\end{table}

\begin{table}
\begin{center}
\begin{tabular}{lrr}
Julian Day   & RV (m\,s$^{-1}$)  & $\sigma$ (m\,s$^{-1}$) \\
\hline
2453253.3359  & -11.0  & 5.4 \\
2453301.2617  & 150.6  & 4.0  \\
2453303.2812  & 162.3  & 3.3 \\
2453332.1953  & 156.5  & 4.7 \\
2453354.1250  & 103.9  & 4.2 \\
2453356.1953  & 123.9  & 4.0  \\
2453429.9609  & 205.7  & 4.0 \\
2453729.0703  & -139.1 & 4.9 \\
2453759.1250  & -181.9 & 4.3 \\
2453761.0703  & -88.8  & 3.8 \\
2453821.0039  & 100.6  & 7.2 \\
2454027.2617  & 123.1  & 5.7 \\
2454111.9882  & -17.8  & 4.2 \\
2454123.0468  & -53.1  & 4.1 \\
2454126.0859  & -140.3 & 5.4 \\
2454209.9531  & -262.1 & 6.6 \\
2454382.3515  & -18.4  & 5.7 \\
2454469.9921  & 163.4  & 4.3 \\
2454471.9921  & 192.7  & 5.9 \\
2454507.0000  & 251.6  & 3.1 \\
2454537.0117  & 194.0  &  3.2 \\
2454755.3476  & -57.3  & 6.9 \\
2454756.2617  & -83.5  & 7.8 \\
2454847.1875  & -107.9 & 4.7 \\
2454880.1015  & -17.2  & 5.9 \\
2454930.9648  & -112.4 & 4.0  \\
2455171.1250  & 182.4  & 6.0 \\
2455248.0156  & 143.0  & 5.6 \\
2455251.0742  & 77.1   & 4.5 \\
2455454.3320  & -32.2  & 5.1 \\
2455554.0664  & -240.0 & 4.7 \\
2455581.1015  & -184.0 & 6.7 \\ 
2455842.3320  & 148.9  & 6.3 \\
2455932.9882  & 152.3  & 5.6 \\
2455962.9453  & 23.9   & 5.8 \\
2456023.9453  & -108.2 & 36.4 \\
2456176.3281  & -27.4  & 4.4 \\
2456257.9960  & -136.1 & 6.6 \\
2456288.0820  & -16.7  & 5.2 \\
2456376.9570  & 90.2   & 4.0 \\
2456377.9960  & 90.1   & 4.0 \\
2456552.2226  & 61.7   & 6.3 \\
\hline
\end{tabular}
\caption{BOAO RV measurements for $\alpha$ Tau.}
\end{center}
\label{boao}
\end{table}

\begin{table}
\begin{center}
\begin{tabular}{lr}
Parameter  & Value  \\
\hline
Period [days]  &   628.96  $\pm$ 0.90 \\
T$_{periastron}$ [JD] & 2451297.0    $\pm$  50.0\\
$K$ [m\,s$^{-1}$] & 142.1 $\pm$  7.2 \\
$e$               & 0.10  $\pm$ 0.05 \\
$\omega$ [deg]    &  287    $\pm$ 29\\
$f(m)$   [M$_\odot$] & (1.84 $\pm$ 0.28) $\times 10^{-7}$ \\
$m$ sin $i$ [$M_{Jup}$]       & 6.47  $\pm$ 0.53\\
$a$ [AU] & 1.46 $\pm$ 0.27\\
\hline
\end{tabular}
\caption{Orbital parameters for the companion to $\alpha$ Tau. 
}
\end{center}
\label{orbitparm}
\end{table}

 
\begin{table}
\begin{center}
\begin{tabular}{lcc}
Quantity  & Period (days) & K (m\,s$^{-1}$)  \\
\hline
RV residuals: Epoch 3    & 521 $\pm$ 11 & 95 $\pm$ 10 \\
RV residuals: Epoch 2    & 526 $\pm$ 82 & 132 $\pm$ 30 \\
RV residuals: Epoch 1-3  & 534.7 $\pm$ 1.7 &  91.2 $\pm$ 9.0 \\
Ca II K S-index          & 521 $\pm$ 10 & N/A \\
H$\alpha$ EW    & 520 $\pm$   23  & N/A \\
H$\alpha$ FWHM  & 526 $\pm$   10  & N/A \\
\hline
\end{tabular}
\caption{Periods derived from RV residuals and activity indicators.}
\end{center}
\label{500}
\end{table}

\clearpage


\end{document}